\documentclass[12pt,aps,notitlepage,superscriptaddress, amsmath,amssymb,amsfonts,floatfix]{revtex4}
\usepackage{amsmath}
\usepackage{graphicx}
\usepackage{epsfig}
\usepackage{amssymb}
\usepackage{xcolor}
\usepackage{tabularx}

\begin{document}

\title{Intermittent chaotic spiking in the van der Pol-FitzHugh-Nagumo system with inertia}
\author{Marzena Ciszak}
	\affiliation{CNR-Istituto Nazionale di Ottica, Via Sansone 1, I-50019 Sesto Fiorentino (FI), Italy.}
	\author{Salvador Balle}
	\affiliation{Department of Ecology and Marine Resources, Institut Mediterrani d’Estudis Avan\c cats, IMEDEA (CSIC--UIB), Balearic Islands, Spain}
	\author{Oreste Piro}
\affiliation{Department of Ecology and Marine Resources, Institut Mediterrani d’Estudis Avan\c cats, IMEDEA (CSIC--UIB), Balearic Islands, Spain}	
\affiliation{Departament de F\'\i sica, Universitat de les Illes Balears, Ctra. de Valldemossa, km 7.5, Palma de Mallorca E-07122, Spain}
		
	\author{Francesco Marino}
	\affiliation{CNR-Istituto Nazionale di Ottica, Via Sansone 1, I-50019 Sesto Fiorentino (FI), Italy.}
	\affiliation{INFN, Sezione di Firenze, Via Sansone 1, I-50019 Sesto Fiorentino (FI), Italy}

\date{\today}

\begin{abstract}

The three-dimensional (3D) Fitzhugh-Nagumo neuron model with inertia was shown to exhibit a chaotic mixed-mode dynamics composed of large-amplitude spikes separated by an irregular number of small-amplitude chaotic oscillations. In contrast to the standard 2D Fitzhugh-Nagumo model driven by noise, the interspike-intervals distribution displays a complex arrangement of sharp peaks related to the unstable periodic orbits of the chaotic attractor. For many ranges of parameters controlling the excitability of the system, we observe that chaotic mixed-mode states consist of lapses of nearly regular spiking interleaved by others of highly irregular one. We explore here the emergence of these structures and show their correspondence to the intermittent transitions to chaos. In fact, the average residence time in the nearly-periodic firing state, obeys the same scaling law ---as a function of the control parameter--- than the one at the onset type I intermittency for dynamical systems in the vicinity of a saddle node bifurcation. We hypothesize that this scenario is also present in a variety of slow-fast neuron models characterized by the coexistence of a two-dimensional fast manifold and a one-dimensional slow one.

\end{abstract}

\maketitle

\section{Introduction}

Complex spike sequences are ubiquitous phenomena in nature and play a relevant role in different aspects of neuroscience, such as in neuronal coding and memory formation \cite{izhikevich2000,mainen95}. In slow-fast neuron models having two or more variables that evolve with very different time scales, firing patterns appear in the form of a mixture of large-amplitude spikes (excitations) and quasiharmonic (subthreshold) oscillations. Such a regime is known as Mixed Mode Oscillations (MMOs) \cite{bronsrev}. Depending on the parameters, the dynamics of mixed mode oscillators can be either periodic or chaotic. In the former case, the periodic spikes are uniformly separated by a regular sequence of small subthreshold oscillations, while in the latter the spikes are delivered at irregular timings and separated by an equally irregular number of subthreshold bumps. 
 
These phenomena \cite{bronsrev} can be originated by various dynamical mechanisms such as, for instance, the quasiperiodic route to chaos on an invariant 2-torus \cite{larter} or the loss of stability of a Shilnikov homoclinic orbit \cite{arneado}. However in slow-fast systems, MMOs are often related to the canard phenomenon in three- or higher-dimensional flows \cite{brons2006,krupa2008,Guck1,Guck2}. 

Canards have been first studied in a two-dimensional (2D) system, namely the van der Pol-FitzHugh-Nagumo equations that were originally conceived as a simple model of neural activity \cite{fhn}. In this system, a limit cycle born from a supercritical Hopf bifurcation experiences an abrupt transition from the small-amplitude regime to large amplitude relaxation oscillations as a control parameter is varied. These sudden changes are known as canard explosions \cite{canard}. The splitting between the time scales makes the transition to occur in an exponentially small range of the control parameters and determines the characteristic shape of the relaxation orbits. Each spike is approximately formed by a sequence of segments of slow motion occurring near the attracting branches of a critical manifold defined by the equilibria of the fast dynamics, separated by rapid switches between these two branches \cite{smale}.

While for a fixed set of parameters 2D systems may either display subthreshold quasi-harmonic oscillation or relaxation oscillations but never a mixture of both, in higher dimensions a clear separation of the two regimes is usually lost and complex MMOs can be observed at different values of the control parameter. Apart from neural systems \cite{alonso89,medvedev04}, examples are also found in chemistry \cite{schmitz,showalter,maselko,brons1991}, plasma physics \cite{mikikian} and optoelectronics \cite{marino2011a}, to name just a few.

In Ref. \cite{marino2007} some of us introduced a minimal model of such dynamics based on a 3D extension of the van der Pol–FitzHugh–Nagumo model where the fast motion includes an inertial term. In this system, that we dubbed \textit{Inertial van der Pol–FitzHugh–Nagumo} (IvdPFN) model, a cascade of period-doubling bifurcations gives rise to periodic and chaotic subthreshold attractors that develop before the canard explosion take place. On each of these attractors, the system is excitable: any fluctuation driving its state variables a critical distance away from the original --now unstable-- fixed point elicits a well-defined response in the form of single relaxation spike. Since the size of the attractor grows as the parameter is increased, part of it exceeds such a critical distance and as a result, an irregular MMO regime sets in. In such a regime, excitable spikes are erratically but deterministically triggered by the chaotically fluctuating background. Experimental evidence of this scenario has been found in semiconductor lasers systems \cite{alnaimee2009,alnaimee2010} and optomechanical resonators \cite{marino2011b,marino2013}.

A distinctive feature of such MMO states is that the distributions of inter-spike intervals ($S$) exhibit a series of sharp peaks at relatively short values of $S$, while at larger values such distributions become exponential and indistinguishable those obtained from a 2D FitzHugh–Nagumo system driven by stochastic forces (see Fig.\ref{figure3}, as wells as \cite{marino2007,alnaimee2009}). The sharp peaks of the distributions are, thus, the fingerprint this type of chaotic spiking and a complete description of its origin and characteristics have not been completely elucidated yet.

Another characteristic of these chaotic MMO systems is the alternance of parameter ranges for which the dynamics is periodic and chaotic. Near the transition between a periodic and a chaotic phase, the system exhibit time intervals of nearly periodic spiking that are interrupted by other intervals of irregular firing. This behavior is reminiscent of intermittency scenario described by Pomeau and Manneville where a nearly periodic system exhibits irregular periods of chaotic dynamics \cite{pomeau}. The type of intermittency has been classified in type I, II or III according to the type of principal bifurcation originating it in paradigmatic models. Such bifurcations are either saddle-node, subcritical Hopf, or inverse period-doubling respectively for each class. However, the three types of intermittency can be diagnosed directly from an intermittent time series by means of three different characteristic universal scaling exponents measuring the divergence of the duration of the chaotic epochs at the transition to the fully chaotic regime.

In this paper, we show that the complex structure of peaks in the distribution for $S$ originates from the intermittent locking of the large amplitude spike train to the small-amplitude chaotic oscillator. From the analysis of the residence times of the system in nearly-periodic epochs and the number of spikes in the time-series, we demonstrate that the transition to the chaotic spiking regime occurs via intermittency of type I. Intermittency of this type has been identified in a variety of dynamical systems, including the Lorenz model and logistic map \cite{eckmann,ott}. We believe that this could be a quite general mechanism for the transition between periodic and irregular mixed-mode states, at least in the context of MMOs via canard-explosions.

The paper is organized as follows. In Sec. II, we introduce the IvdPFN model and we present an overview of the system dynamics with particular attention to the MMO regimes. In Sec. III we analyze the inter-spike intervals of MMO states and their probability distributions in the small range of parameters around the transition between periodic to chaotic spiking. In Sec. IV we characterize the average residence times as a function of the control parameter and we show that such quantities obey the characteristic scaling law of type-I intermittency. We finally show in Sec. V that the recurrence map for $S$ displays the tangency typically associated with the type I intermittent behavior, as also confirmed by the direct analysis of the Poincar\'e map of the system reported in Sec. VI. Conclusions and future perspectives are presented in Sec. IV.
\begin{figure}
\begin{center}
\includegraphics*[width=1.0\columnwidth]{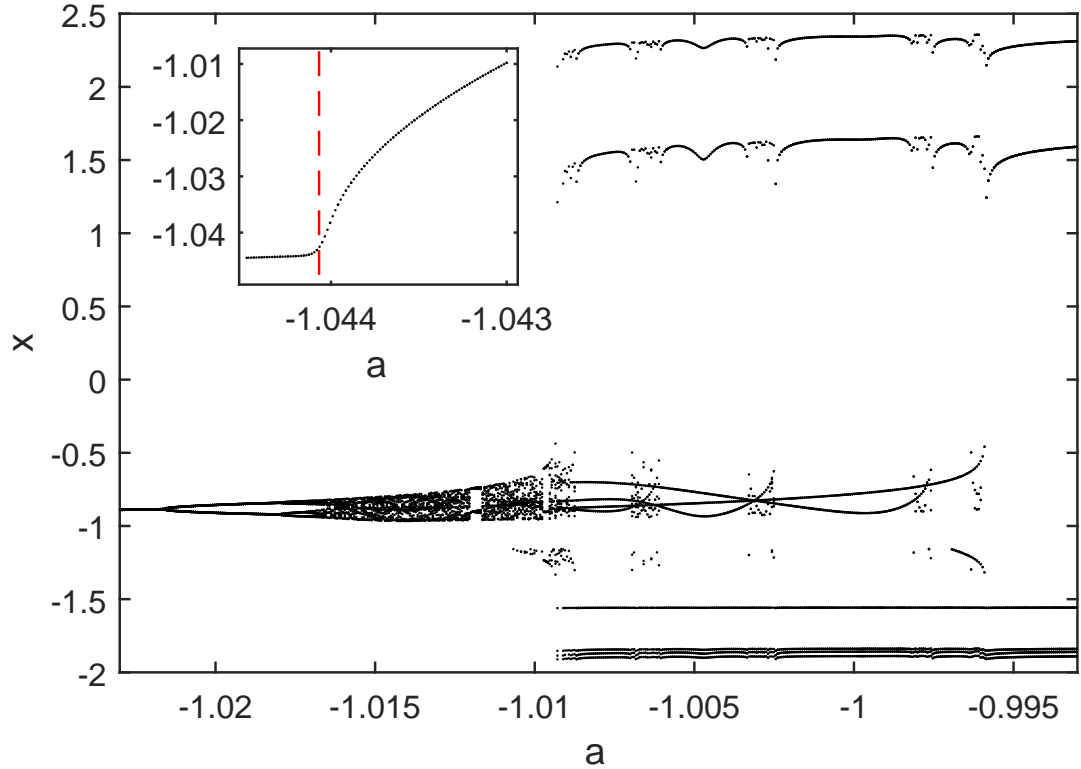}
\end{center} 
\caption{Bifurcation diagram of the Poincar\'e map of the attractor of the IvdPFN model as the parameter $a$ is varied. The inset shows the onset of the Hopf bifurcation at $a=-\sqrt {1+\varepsilon k}$ The plot points represent the local maxima of $x(t)$ coincident with the crossings at $z(t)=0$ for $\varepsilon=0.03$ and $k=3$.}
\label{figure1}
\end{figure}
\begin{figure}
\begin{center}
\includegraphics*[width=1.0\columnwidth]{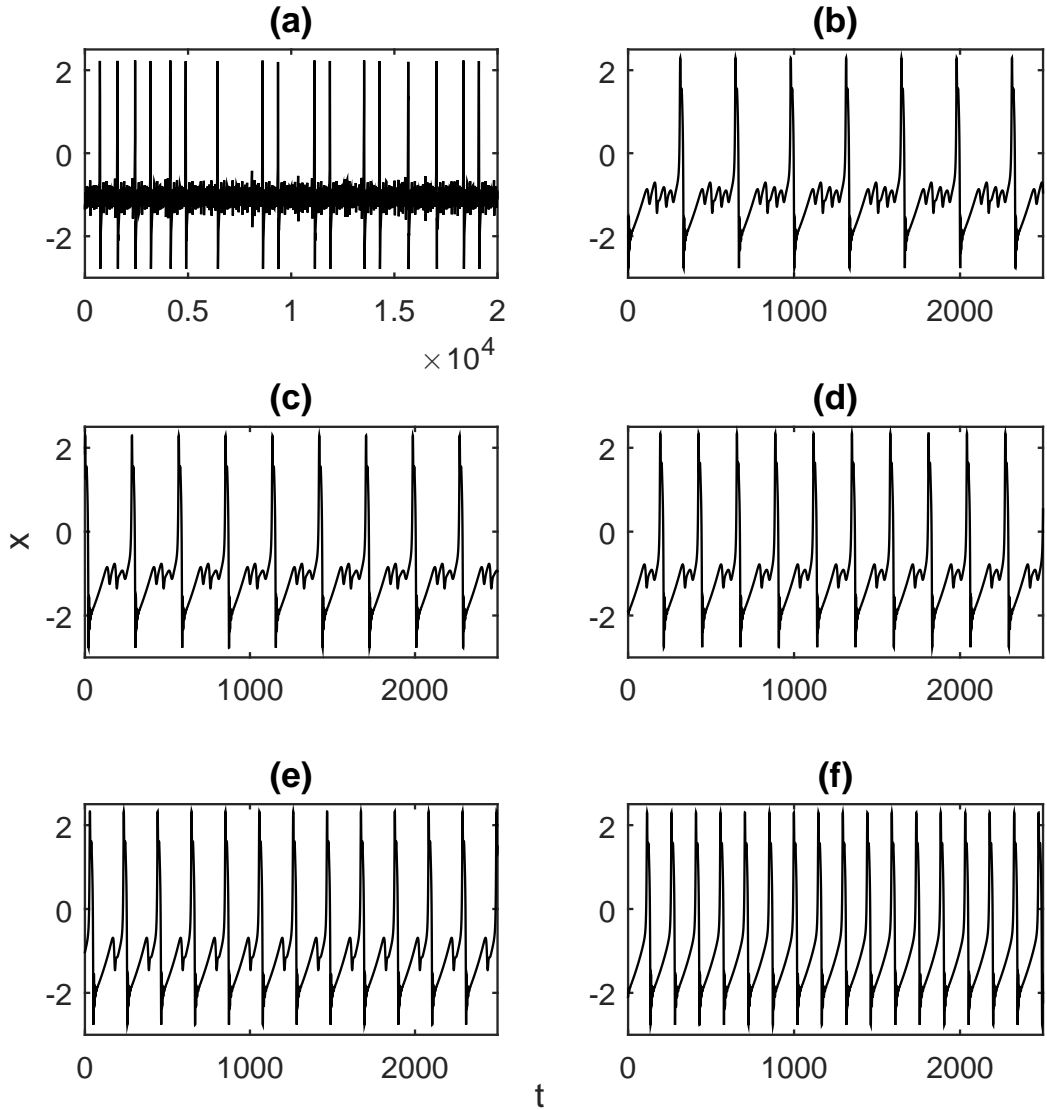}
\end{center} 
\caption{Mixed mode oscillation regimes (a) chaotic at $a=-1.009$, (b) $1^4$ at $a=-1.0075$, (c) $1^3$ at $a=-1.005$, (d) $1^2$ at $a=-1$, (e) $1^1$ at $a=-0.997$ and (f) $1^0$ at $a=-0.994$.} 
\label{figure2}
\end{figure}

\begin{figure}
\begin{center}
\includegraphics*[width=1.0\columnwidth]{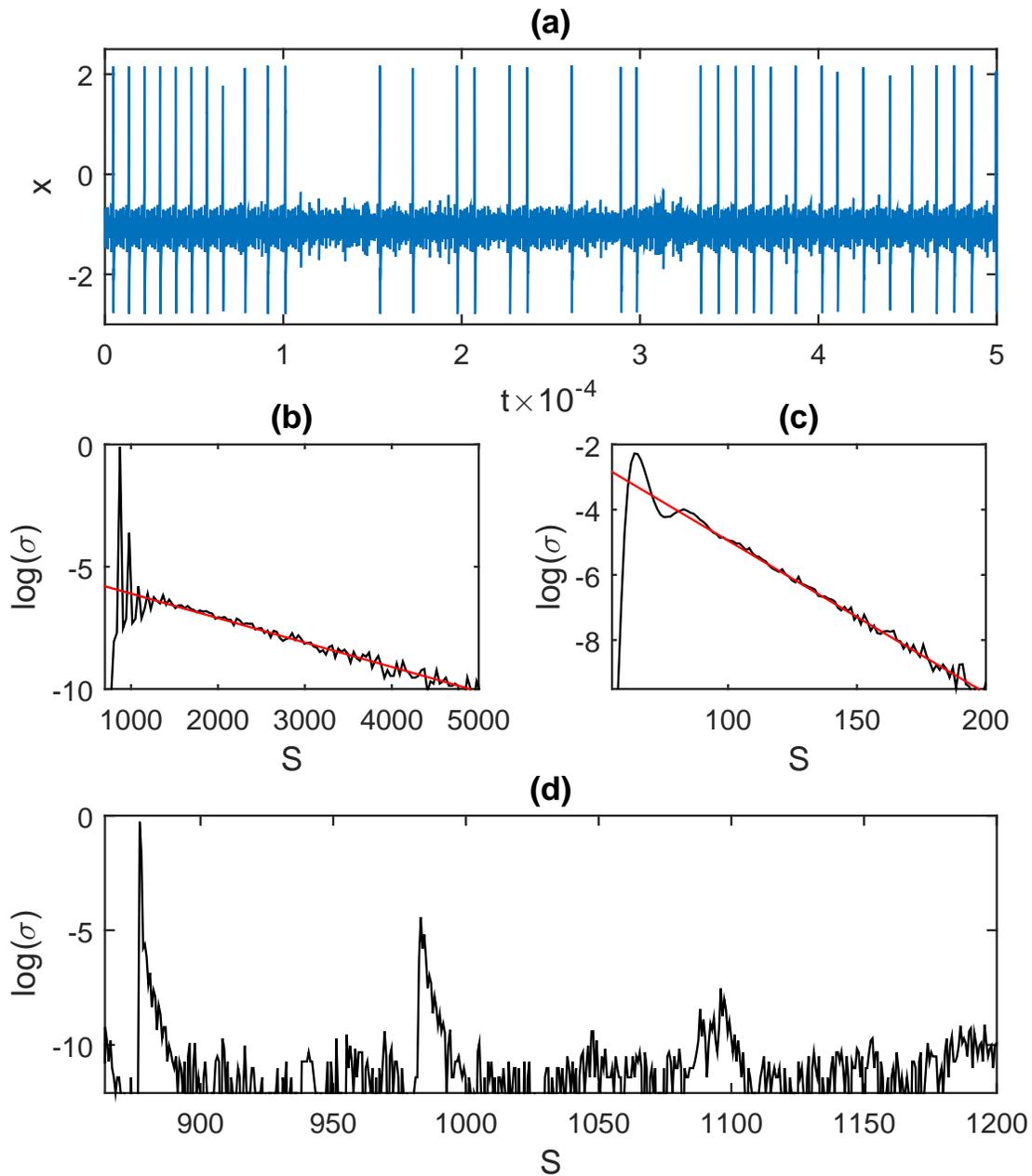}
\end{center} 
\caption{(a) Time series for the variable $x$ at $a=-1.0092504965128$. Distributions of inter-spike intervals $S$ for (b) $a=-1.00927$ and (c) $a=-1.0092504965128$. (d) The blow-up of the distribution from (c).} 
\label{figure3}
\end{figure}
\begin{figure}
\begin{center}
\includegraphics*[width=1.0\columnwidth]{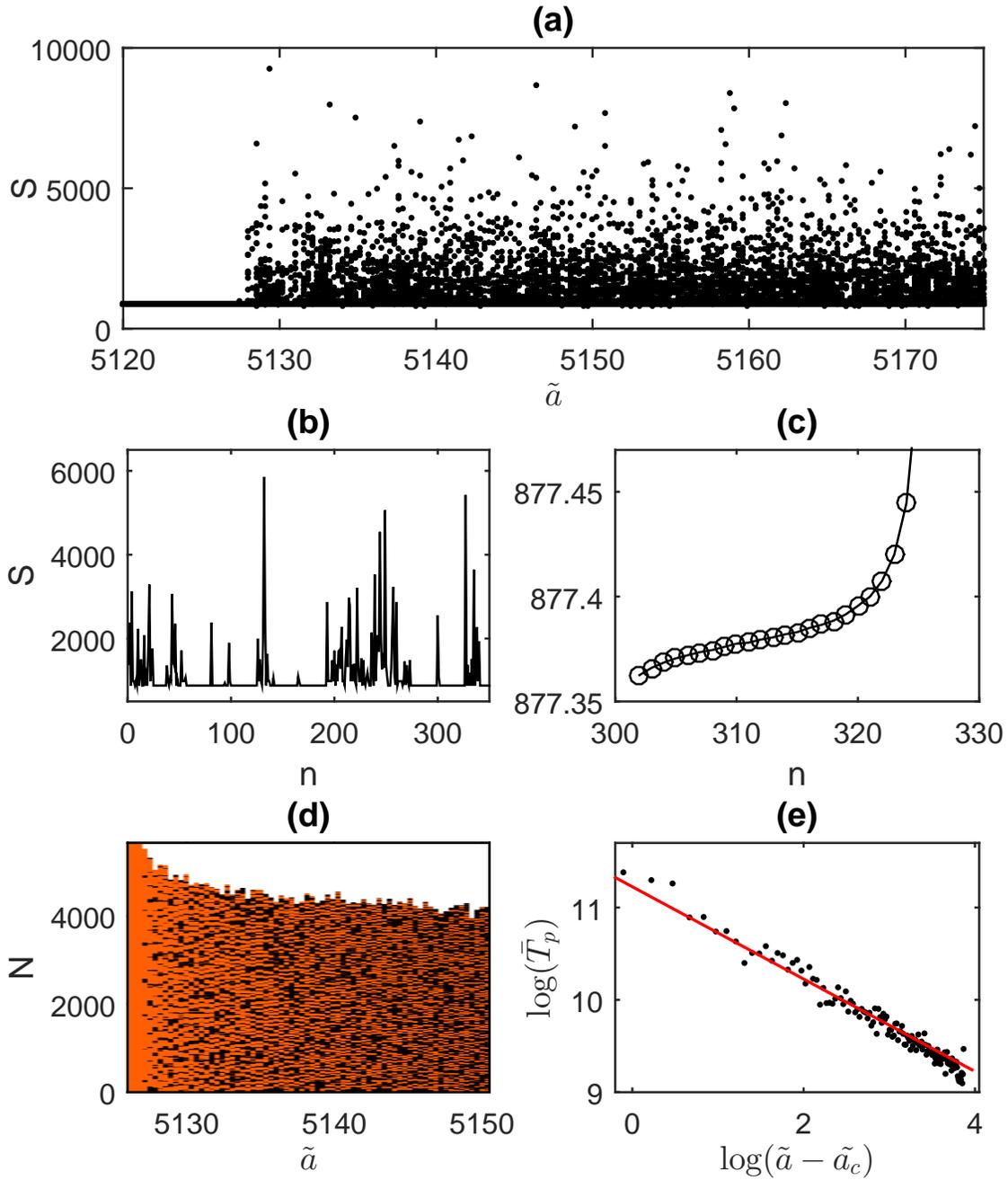}
\end{center} 
\caption{(a) ``Bifurcation diagram'' for $S$ as parameter $a$ is varied. Abscissas of the points along the vertical lines are the successive values of $S$ in an orbit for the corresponding value of the parameter $a$  (b) Selected sequence of successive $S$ for a=-1.0092504965175. (c) Blow-up of a small region from (b). (d) Total number of spikes $N$ contained in a fixed time interval ($\Delta t=5\times 10^6$) for different decreasing values of the parameter $a$ (increasing $\tilde {a}$). In each bin bar we color-discriminate the sub-sequences with almost constant $S\approx \Sigma_1$ (orange segments) from the sub-sequences of substantially different $S's$ (black segments). The number of spikes in each sub-sequence is proportional to the length of the corresponding colored segment. (e) Log-log plot of the mean residence times $\bar{T}_p$ versus $\tilde{a}$.}
\label{figure4}
\end{figure}
\begin{figure}
\begin{center}
\includegraphics*[width=1.\columnwidth]{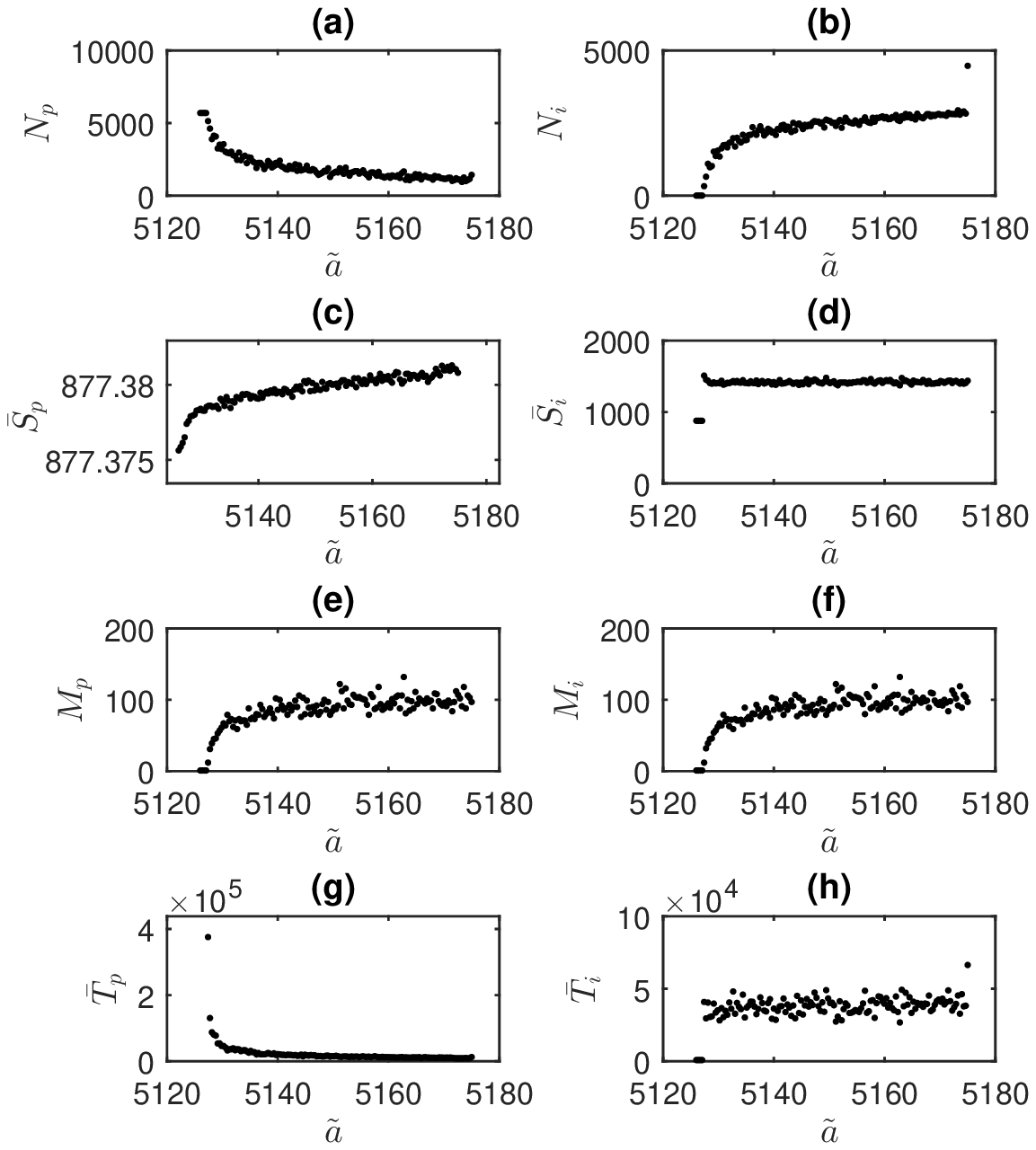}
\end{center} 
\caption{(a) Number $N_p$ of nearly constant inter-spike intervals $S\approx \Sigma_1$. (b) Number $N_{i}$ of irregularly spaced spikes $S_i$. (c) Mean of the nearly periodic $S$: $\bar{S}_p$. (d)  Mean of the irregular $S$: $\bar{S}_i$.
(e) Number of the periodic residence time epochs $M_p$. (f) Number of the irregular residence time epochs $M_{i}$. (g) Mean residence time of nearly periodic $S$: $\bar{T}_p$. (h) Mean residence time of irregular $S$ $\bar{T}_i$. All curves are drawn versus the control parameter $\tilde{a}$. 
}
\label{figure5}
\end{figure}

\section{Model equations}
The IvdPFN system under consideration in this paper is a modification of the  van der Pol-Fitzugh-Nagumo model (vdPFN) with the addition of an inertial term in the dynamics of the fast variable $x$. The defining equations of the IvdPFN \cite{marino2007} read:
\begin{eqnarray}
k\ddot{x}+\dot{x}&=&y+x-x^3/3  \label{eq1} \\
\dot{y}&=&-\epsilon (x-a) \label{eq1b}
\end{eqnarray}
where in the singular limit of  $k\rightarrow 0$ we recover the original vdpFN. In this original case, $\epsilon$ is a small parameter that separates two different time scales and the external control parameter $a$ governs the transition from excitability to periodic self-pulsing via a Hopf bifurcation (at $a=\pm 1$) followed, as $a$ increases, by a canard explosion. In the modified case, instead, the Hopf bifurcation occurs at slightly smaller values of $a$ and it is followed   by a sequence of period doubling bifurcations leading to a small chaotic attractor as $a$ increases. This attractor keeps growing with $a$ and eventually part of it crosses the excitability threshold sporadically delivering spikes as a result.

A better understanding of this phenomenon can be achieved by rewriting Eqs.(\ref{eq1}-\ref{eq1b}) as a system of three first order differential equations through the addition of the extra dynamical variable $z$:

\begin{eqnarray}
k\dot{z}&=&-z+y+x-x^3/3 \label{eq1ab}\\
\dot{y}&=&-\epsilon (x-a) \label{eq2ab}\\
\dot{x}&=&z \label{eq3ab}
\end{eqnarray}
where $\epsilon$ and $a$ play the  same r\^ole as before.  The extra
parameter $k$ determines the relative importance of the inertial motion and its time scale. We focus our attention on the regime $k  \gtrsim  1 \gg \epsilon $ where the fast relaxation and inertia are comparably important in the fast dynamics and $k>1$ corresponds to inertial regime of underdamped oscillations. Throughout this paper $\epsilon$ will be kept fixed at $\epsilon=0.03$ and $k$ at $k=3$.  

The system of Eqs.(\ref{eq1ab})-(\ref{eq3ab}) has a fixed point at $(x,y,z)=(a,a^3/3-a,0)$. Since there is a complete reflection symmetry in $a$ we will investigate only the dynamical behavior of the system for the range of negative $a$ and in particular near $a\approx -1$. Within this range, the fixed point undergoes Hopf bifurcations at $a=- \sqrt{1+\epsilon k}$ giving rise to small amplitude limit cycles as it is illustrated in the inset of Fig. \ref{figure1}.  As $a$ increases beyond the Hopf point, the limit cycle experiences a cascade of period doubling bifurcations leading to a chaotic attractor which amplitude is illustrated in Fig. \ref{figure1}. Notice also  that the small attractor grows in size roughly monotonously with $a$. The growing attractor starts intersecting the canard manifold at $a\approx -1.01$ and chaotic spiking arises as a consequence of this intersection. This gives rise to the appearance of mixed-mode oscillations composed by irregularly spaced large amplitude spikes interleaved by intervals of small chaotic oscillations as shown in Fig. \ref{figure2} (a).

As $a$ is further increased past the canard explosion point we find different $a$-windows where periodic and chaotic spiking are observed in some analogy to what happens for the logistic map deeply in the chaotic regime. Each periodic window originates in a saddle-node (or tangent) bifurcation giving birth to a mixed mode closed orbit composed by segments of large excursions in the state space interspersed by much smaller spiral segments in the vicinity of the unstable fixed point of the system. Plotted as a time-series these closed orbits appear as a periodic sequence of large spikes sandwiched by complex patterns of sub-threshold oscillations near the unstable rest state. In turn, these periodic closed orbits undergo sequences of period doubling bifurcations leading to chaotic mixed-mode spiking. We label the dynamics of the mixed-mode periodically spiking windows  by a sequence of symbols $L^S$ each standing for number $L$ of large amplitude spikes followed by number $S$ of small sub threshold oscillations. The ratio $L/S$ of the number of spikes to the number of sub-threshold oscillations  are often referred as {\it winding numbers}. For the time scales that we consider in this paper, $L$ is always equal to 1. Various cases of $1^S$ dynamics are shown in Fig. \ref{figure2} (b-f). 
\begin{figure}
\begin{center}
\includegraphics*[width=1.0\columnwidth]{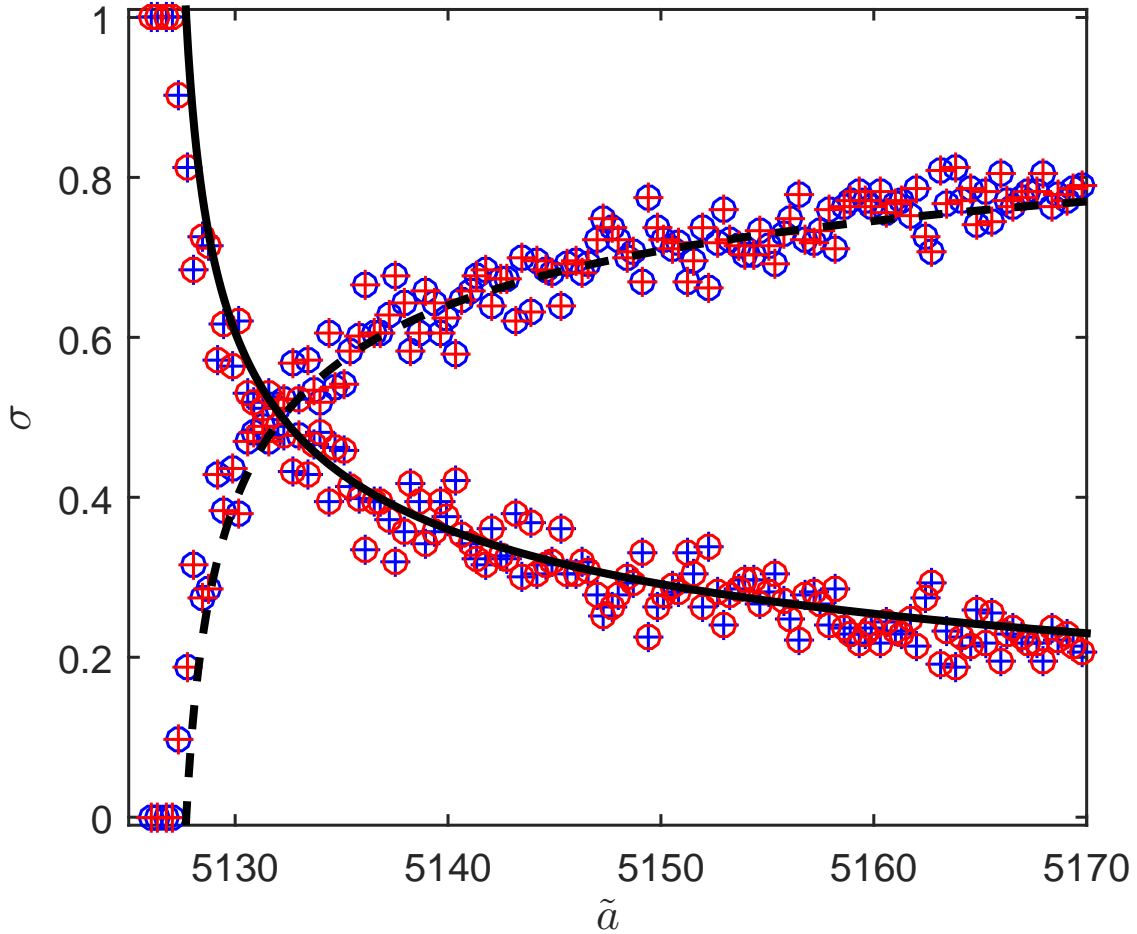}
\end{center} 
\caption{Probabilities $\sigma $ calculated numerically versus control parameter $\tilde{a}$ are marked as follows: $\sigma_{p}^{T}$ - blue +, $\sigma_{p}^{S}$ - red o, $\sigma_{i}^{T}$ - blue o, $\sigma_{i}^{S}$ - red +. Black lines show the fitted functions. Solid line is for $\sigma_{p}^{T,S}$ and dashed line is for $\sigma_{i}^{T,S}$, all defined in Eq. \ref{eq6}.}
\label{figure6}
\end{figure}
\begin{figure}
\begin{center}
\includegraphics*[width=1.0\columnwidth]{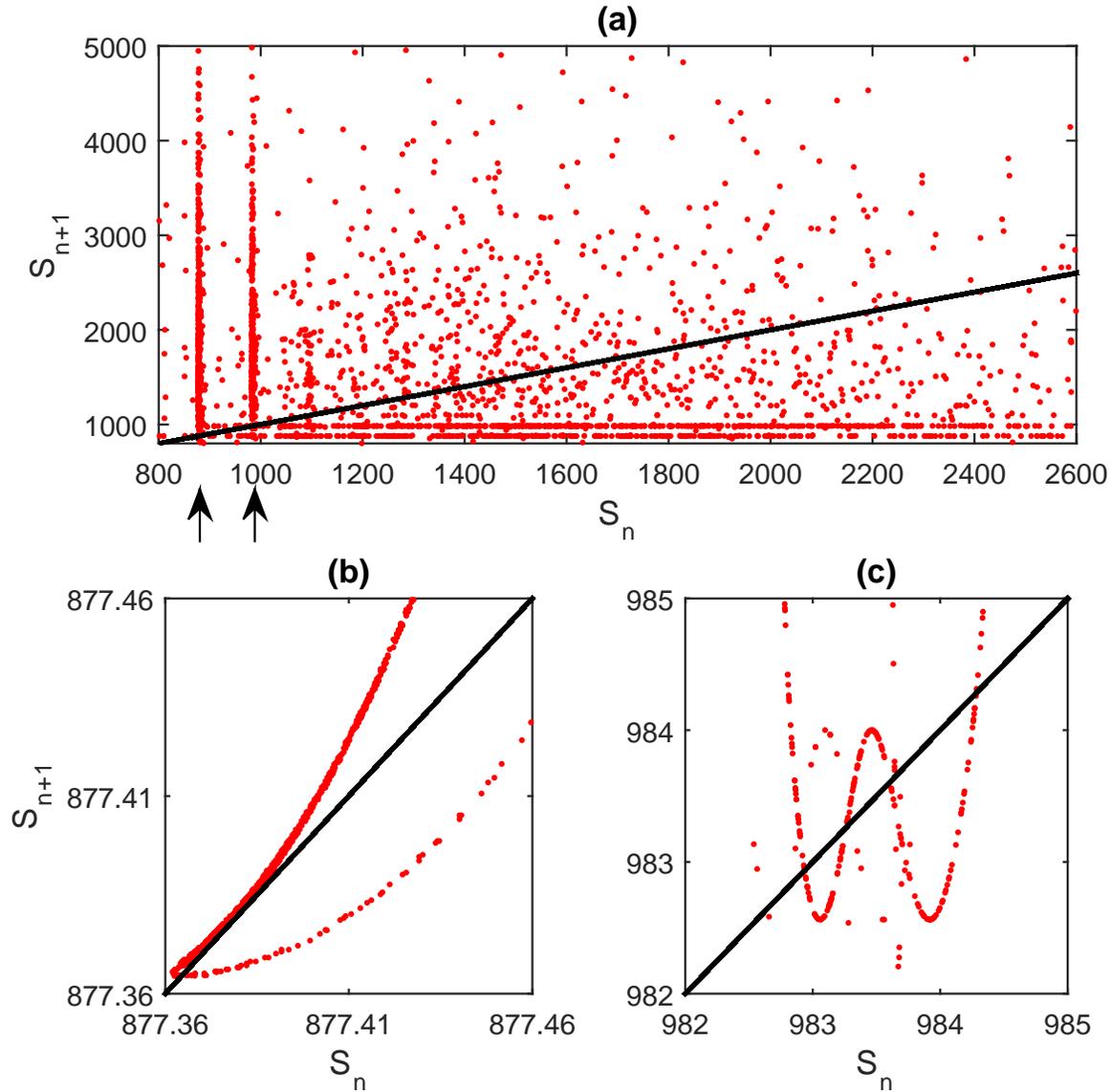}
\end{center} 
\caption{(a) Recurrence plot for $S_n$ at $a=-1.0092504965128$. Two vertical arrows mark the blow-up regions in the map shown respectively in (b) and (c). The black line is $S_{n+1}=S_n$.
}
\label{figure7}
\end{figure}

\begin{figure}
\begin{center}
\includegraphics*[width=1.0\columnwidth]{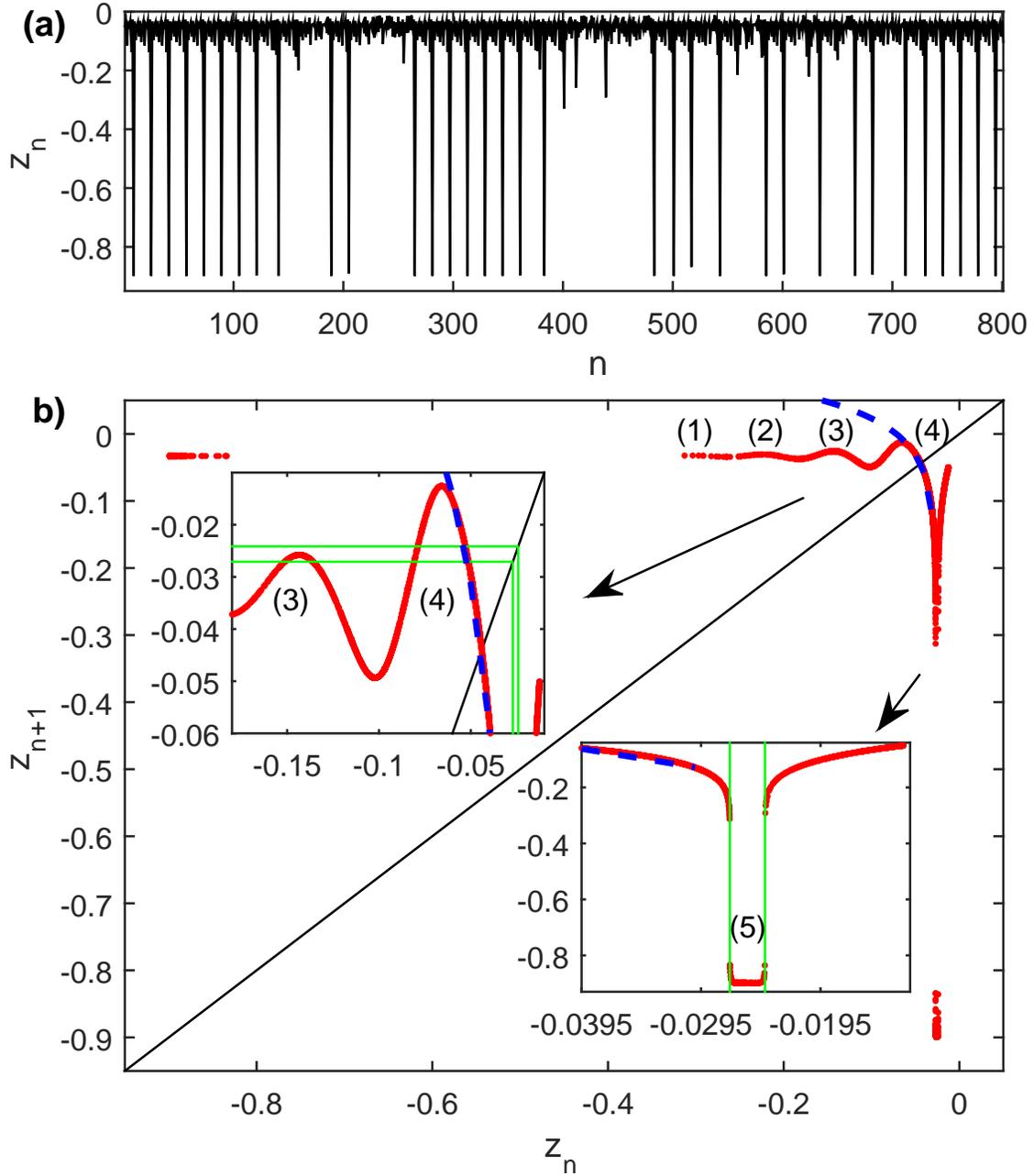}
\end{center} 
\caption{(a) Evolution of $z_n$ obtained from a Poincar\'e map at $a=-1.0092504965128$. (b) Poincar\'e map (red dots), four oscillations are marked with number (1-4) and a deep well with number (5); logarithmic function (blue dashed line); two straight green lines mark the region on (3) and (4) which when are touched during the iteration of a map give in the subsequent iteration the value in a well (5), giving rise to a large spike. The black line is $z_{n+1}=z_n$.}
\label{figure8}
\end{figure}
\begin{figure}
\begin{center}
\includegraphics*[width=1.0\columnwidth]{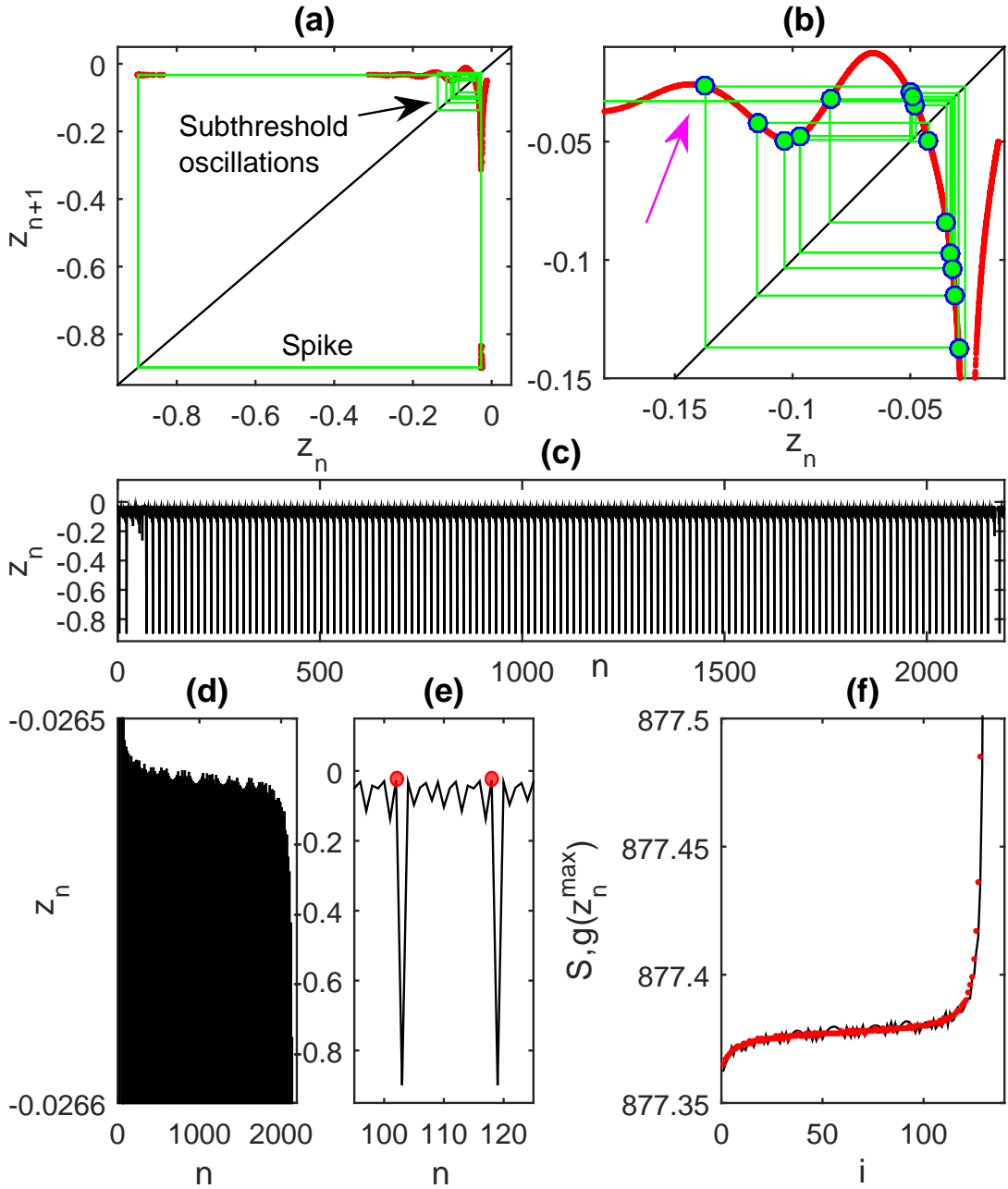}
\end{center} 
\caption{(a-b) Poincar\'e map for a system in chaotic regime at $a=-1.0092504965128$ (red color) and $z_{n+1}=z_n$ (black color). Green lines show the iteration of a map for a single $S$ belonging to the group $\Sigma_1$, the filled circles are the map for a system in periodic regime. Arrow in magenta marks the segment of the map function considered in (d-e). (c) Evolution of $z_n$ with a long periodic intermittent epoch. (d) The blow-up of the maxima of the series shown in (c). (d) Details of a single $S$, where the maxima of the series are marked with red circles. (e) Evolution of $S$ (red color) and the function $g(z_{n}^{max})$ (black color) for the series shown in (c).}
\label{figure9}
\end{figure}
\begin{figure}
\begin{center}
\includegraphics*[width=1.0\columnwidth]{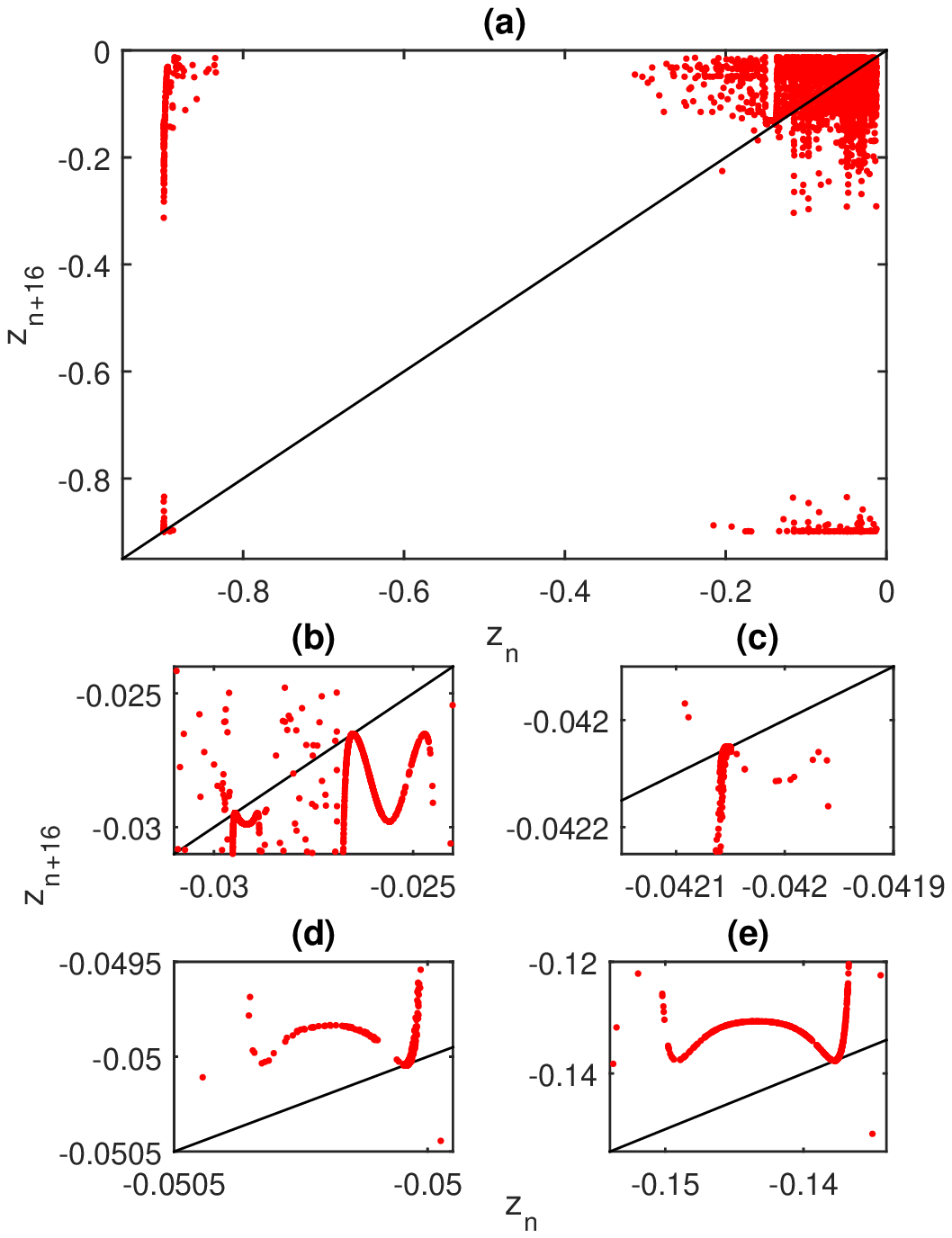}
\end{center} 
\caption{(a) Poincar\'e map $z_{n+16}$ vs $z_n$ at $a=-1.0092504965128$ (red color). (b-e) the blow-up of selected regions in (a). The black line is $z_{n+1}=z_n$.}
\label{figure10}
\end{figure}
\section{Probability distributions of inter-spike intervals}

An interesting analogy between the effects of the small chaotic attractor in the dynamics of autonomous 3D IvdPFN and that of a small random force driving a noisy 2D vdPFN was pointed out in \cite{marino2007}. In fact, for certain values of the parameter $a$ the probability distributions $\sigma(S) $ of interspike intervals $S$ produced by the IvdPFN and the corresponding one of the random spikes delivered by the noise driven two-dimensional FitzHugh-Nagumo model do not differ significantly in the limit of large $S$ values. Both $\sigma(S)$ distributions are fitted accurately by an exponential function as it is shown for comparison in  Figs. \ref{figure3} (b) and \ref{figure3} (c). However, in the opposite range of small $S$ values just above the minimum possible dictated by the refractory time-scale, we can appreciate the presence of a complex structure of unstable periodic orbits which cannot appear in the noisy vdpFN. The signature of these orbits is displayed in Fig. \ref{figure3} (d) where we show a blow-up of the range of $a$ where the distribution of Fig. \ref{figure3} (c) shows peaks. The blow-up reveals that each peak has a smoothly decaying tail that corresponds to a gradual increase of $S$ in the neighborhood of a given unstable periodic orbit. We will return to this point with further details in the next section discussion referred to Fig. \ref{figure4} (c). It should be noticed that as the parameter $a$ increases the system develops an increasingly stronger chaotic regime where unstable periodic orbits proliferate eventually forming a dense set. As a consequence of this proliferation, the corresponding structure of peaks in the distribution of interspike intervals becomes more and more complex until finally merging into a dominating exponential form.

Let us now analyze in detail the small range of parameters where the abrupt transition from periodic to chaotic spiking is observed. This range is shown in Fig. \ref{figure4} (a) in the form of a ``bifurcation diagram'' for the sequences $S_n$ generated for each value $a$. There, the symbol $\tilde{a}$ is defined as the last four decimals of the corresponding value of $a$, e.g.  $a=-1.009250496(\tilde{a})$. In the single temporal trace of $S_n$ shown in Fig. \ref{figure4} (b) a sequence of apparently periodic spikes is interleaved with segments of irregularly spaced spikes separated by substantially longer interspike intervals $S$'s. This behavior is strongly reminiscent of chaotic intermittency. All $S$ belonging to a given peak in the distribution have similar lengths and contain the same number of subthreshold oscillations. There are however small differences in the subsequent lengths of $S$ as shown in Fig. \ref{figure4} (c), that are much smaller than the duration of a single subthreshold oscillation. These small differences lead to the gradual increase of $S$ ending up in the transition to irregular bursting with longer $S$. Such increase of $S$ contributes to the appearance of the characteristic peaks in the distribution of $S$. Fig. \ref{figure4} (b) shows these epochs of different duration during which $S$ smoothly increases, occurring at different times and separated by bursts of chaotic spiking.

\section{Residence times}
In the previous section we have shown the occurrence of epochs of nearly periodic spiking interrupted by others of chaotic spiking. In order to characterize this phenomenon we will focus on the duration of these nearly periodic epochs which we will call \textit{residence times}. The residence time $T_p$ ---where $p$ stands for periodic--- is defined as the length of consecutive sub-sequence of $S$'s that fall within a given range. In our case we consider, for example, the interval $\Sigma_1=\{877.36 \leq S \leq 877.4\}$, corresponding to $S$'s in the first and largest peak of the distribution displayed in Fig. \ref{figure3} (d). The shortest value in $\Sigma_1$ corresponds to the stable orbit occurring at given value of $a$ in the periodic regime, just below the point where the transition to intermittent behavior occurs. On the other hand, values of $S$ outside this interval are considered belonging to the irregular bursting phase composed by a successive sub-sequence of chaotic spikes. 

It is worth to note that the second large peak of the distribution, corresponding to the range $\Sigma_2=\{983 \leq S \leq 1000\}$ of interspike intervals, is much smaller than that of values of $S$ in $\Sigma_1$. Hence  $S \in \Sigma_2$ are significantly less probable than  $S\in \Sigma_1$. This is due to the fact that the orbits responsible for the $\Sigma_2$ trains are significantly more unstable than those  corresponding to $\Sigma_1$. Therefore, we can safely focus only on $\Sigma_1$ for the computation of $T_p$ and consider $\Sigma_2$ as part of the irregular phase. Notice in passing that the signals corresponding the set $\Sigma_2$ contain one extra sub-threshold oscillation compared to those of the set $\Sigma_1$. 

For each value of $a$ starting from one in the periodic spiking regime, we count the total number $N$ of spikes delivered during a fixed observational time. It turns out that $N$ decreases as $a$ increases because the system tends to remain longer in the subthreshold oscillatory state hereby contributing to longer values of  $S$  (see Fig. \ref{figure4} (d)). The decrease of $N$ is logarithmic for a sufficiently long observational time windows. This means that the residence time in the nearly periodic state $T_p$ decreases as $a$ increases. Computing the average of this quantity over many realizations we find that the following scaling law:

\begin{eqnarray}
\bar{T}_p \propto (a-a_c)^{-\gamma} 
\label{eq2}
\end{eqnarray}
with the critical exponent $\gamma =\frac{1}{2}$ (see Fig. \ref{figure4} (e)). This coincides with the scaling law obeyed at the transition to chaos via intermittency of type I \cite{pomeau}. 

To provide further evidence of the connection to this type of intermittency, let us now study the relation between residence times and the dynamics involved in the interspike sequences $S_n$. Notice first that the following conservation equations must hold: 
\begin{eqnarray}
M_p\bar{T}_p+M_{i}\bar{T}_i=t \nonumber \\
N_p\bar{S}_p+N_{i}\bar{S}_i=t.
\label{eq3}
\end{eqnarray}
Here $p$ stands for {\it periodic} and $i$ for {\it irregular} and, letting the index $k$ take either the name $p$ or $i$, $M_k$ is respectively the number of nearly periodic or irregular residence times, $N_k$ is the number of nearly periodic (respectively irregular) interspike intervals $S$, $\bar{T}_k$ is the mean duration time of the nearly periodic or irregular epochs and $\bar{S}_k$ is the mean value of $S$ within the nearly periodic (respectively irregular) phases. Finally $t$ is the total time duration of numerical simulations. In Fig. \ref{figure5} (a-h) we plot the numerical estimations for the quantities defined in Eq. \ref{eq3}. Notice that we should normally have $M_p \approx M_i=M$ because to each regular sequence should follows an irregular one (compare Figs. \ref{figure5} (e) and (f)). However this is not true for the extreme cases of a purely periodic regime where we have $M_p=1$ and $M_i=0$, and of a purely chaotic spiking sequence for which $M_p=0$ and $M_i=1$. On the contrary, no such trivial relations exist neither for the corresponding number of spikes nor for interspike intervals $S$'s contained in each of the phases, i.e. $N_p \neq N_i$ as we can see by comparing Figs. \ref{figure5} (a) and (b). Nevertheless, they are still subject to  the obvious constraint $N_p + N_i=N$, where $N$ is the total number of spikes in the simulation. 

Dividing  both Eqs. (\ref{eq3}) by $t$ we get:
\begin{eqnarray}
\sigma_p^{T}+\sigma_i^{T}=\sigma_p^{S}+\sigma_i^{S}=1
\label{eq4}
\end{eqnarray}
for the corresponding probabilities.

In Fig. \ref{figure6} we plot these probabilities revealing that $\sigma_{i}^{T}=\sigma_{i}^{S}$ and $\sigma_{p}^{T}=\sigma_{p}^{S}$. The functions that best fit these numerical data (drawn in solid and dashed black lines in the figure) are:

\begin{eqnarray}
\sigma_{p}^{T,S}&=&\log\left[1+B(a-a_c)\right]^A(a-a_c)^{-\frac{1}{2}} \\
\sigma_{i}^{T,S}&=&1-\sigma_{p}^{T,S}
\label{eq6}
\end{eqnarray}
where the coefficients are $A=0.18$ and $B=100$.

The former analysis shows the equivalence of the approach based on residence times and the one considering only the number of interspike intervals $S$ to characterize intermittency in chaotically spiking systems. In fact we can easily infer that

\begin{eqnarray}
\bar{T}_p \propto N_p \propto (a-a_c)^{-\frac{1}{2}} 
\label{eq5}
\end{eqnarray}

In other words, the same intermittency scaling law displayed by the residence times distribution can be directly estimated by counting the number of nearly periodic intervals $S$ in the time series. 

\section{Recurrence map for S}

The statistics of the residence times is a consequence of a phenomenon of gradual increase of $S$ in the subsequences belonging to the nearly periodic or laminar epochs. Let us then investigate in further detail the dynamical behavior of this quantity by means of the so called return or recurrence map constructed from the sequences $S_n$. 
Fig. \ref{figure7} (a) shows a plot of $S_{n+1}$ versus $S_{n}$ (red dots) for $a=-1.009250496517535$. The plot looks like an apparently irregular cloud of points with some distinguishable features: dots appear to have the tendency to accumulate near some horizontal and vertical lines are correlated with the peaks in the distributions shown in Fig. \ref{figure3}. These lines are also fingerprints of intermittency in the return map. Notice, the most prominent of these accumulations occur for values of $S_n\in \Sigma_1$ and $S_n\in \Sigma_2$ for the vertical ones and of $S_{n+1}\in \Sigma_1$ and $S_{n+1}\in \Sigma_2$ for the horizontal ones. To better understand the formation of these lines as a consequence of intermittency, we should zoom into the intersection of these accumulations with the diagonal line
$S_{n+1}=S_n$ shown in black in the picture.  Fig. \ref{figure7} (b) shows a blow-up of this intersection for the case of $S_n\approx S_{n+1}\in\Sigma_1$ revealing a nicely organized pattern emerging from the apparently chaotic cloud: dots lay on a smooth curve almost tangent to the diagonal. Dots on these curve obviously correspond to nearly periodic sequences where $S_n$ smoothly increases with $n$. The existence of such approximate tangency is the defining characteristic of type I intermittent dynamics where orbits take many iterates to go through the narrow tunnel between the map function and the nearly tangent $S_{n+1}=S_n$ line. The approach to tangency is also an indication of the nearness of  a saddle-node (also named \textit{tangent} for obvious reasons) bifurcation for the $S_n$ dynamics, giving rise to a stable periodic sequence. While through the above mentioned tunnel, the orbit seems to be approximately periodic but once exiting it, a segment of chaotic motion dictated by the complex form of the whole map sets in giving rise to the irregular components that configures the picture in Fig. \ref{figure7} (a). A different type of pattern shown in Fig. \ref{figure7} (c), appears in the vicinity of the intersection of the lines corresponding to second peak $\Sigma_2$. While all the dots plotted in this inset represent repetitive sequences of inter-spike intervals closely equal contained in $\Sigma_2$, they are organized in a much more complex structure that reflect the complicated geometry of the whole chaotic attractor. 

Before concluding this section, we should remark that the recurrence map of the inter-spike intervals only help to characterize some aspects of the time series of the spiking. It cannot be considered as a faithful description of the dynamics not even in an approximate sense. Such a description can only be achieved by studying the successive intersection of the actual orbits of the dynamical system with an appropriate (two-dimensional in this case) hyper-surface in the phase space. In the next section we perform this study which provide a more direct evidence for intermittency.

\section{Poincar\'e map}

In order to provide a more rigorous dynamic evidence for the intermittency described previously we now focus  our attention on the Poincar\'e map of the IvdPFN system defined by Eqs.
(\ref{eq1ab}-\ref{eq3ab}). While the actual Poincar\'e map is two-dimensional, we will consider a 1D approximation that turns out to be accurate enough for our purposes. Let $z_n$ be the values of the variable $z$ at the successive intersections of a trajectory with the surface $x=-1$ in the 3D phase space. The sequence $z_n$ is plotted as a function of $n$ in Fig. \ref{figure8} (a). Interestingly, this plot reveals a spiking behavior reminiscent of that of the continuous variable, with sections of different lengths of regularly spaced spikes separated by sections of non-regular spiking. However, the recurrence plot $z_{n+1}$ versus $z_{n}$ shows a well organized 1D structure as the one shown in Fig. \ref{figure8} (b).

From the plot, we can infer the existence of a well defined mapping function $z_{n+1}=f(z_{n})$ composed by 4 oscillations of increasing amplitude marked  (1), (2), (3) and (4) in the figure and ending in a deep well marked as (5). It is worth to mention that the edge of the well (5) has a logarithmic shape plotted as a blue dashed curve. This shape is connected to the logarithmic dependence of $N_p$ on the parameter $a$, described in the previous sections.

The iterations of the map eventually hit the segments of the curve marked as (3) and (4)  inside the region delimited with two green horizontal lines and then fall down into the well (5). These events originate the large excursion giving rise to the formation of a spike in $z_{n}$. On the the contrary, when iterations fall into other parts of the curve (1-4), they remain in the subthreshold oscillatory state. 

Let us now set our attention on the group of spikes separated by interspike intervals on the range $\Sigma_1$ responsible of the largest peak in the distribution of $S$. In Fig. \ref{figure9} (a) we show the iteration corresponding to a given $S$ in the $\Sigma_1$ group range which is composed by sequences  of small subthreshold oscillations separated by large amplitude spikes approximately separated by $S$. The shortest $S$ in $\Sigma_1$ corresponds to the periodic orbit observed at $a$ just below the transition to intermittency. The mapping for this periodic orbit is indicated with filled circles in Fig. \ref{figure9} (b). 

In the intermittent phase, the mapping of  the nearly periodic epochs of the original variable $x$ appear as the perfectly periodic epochs as the one in Fig. \ref{figure9} (d), because between two successive spikes there is a fixed number of iterations corresponding to a fixed number of subthreshold oscillations in $x$. In other words, in the map setting the information about the gradual increase of $S$ during the nearly periodic epochs is hidden. However, gradual differences in  amplitude of the subthreshold oscillations can be noticed in the mapping. For example, \ref{figure9} (d) shows the evolution of the largest maxima of $z_n$ in each $S$ which after some linear scaling coincide with the gradual increase of $S$ observed for the continuous-variable system shown in Fig.  \ref{figure4} (c). The relation between these two quantities can be estimated numerically to be:
$$S=g(z_n^{max})=-1900z_n^{max}+826.966$$
where $z_n^{max}$ are the largest maxima of $z_n$. Every time that the system reaches $z_n^{max}$, indicated by red circles in Fig. \ref{figure9} (e) the following iteration ends up in the deep well (5). To stress the correlation between $S$ and $z_n^{max}$ we plot both $S$ and $g(z_n^{max})$ in Fig. \ref{figure9} (f). The region of the map function, from which the transition to the large spike occurs is therefore responsible for the intermittent behavior in which the slow departure from the nearly periodic to chaotic bursting occurs.

The spikes from $\Sigma_1$ have an iteration period of $16$. Hence in order to examine the subsequent dependence of specific extrema of subthreshold oscillations in $\Sigma_1$ we plot $z_{n+16}$ vs $z_n$ in Fig. \ref{figure10} (a). Notice the arrangement of the iterations in curves tangent to the line $z_{n+1}=z_n$ (Figs. \ref{figure10} (b-e)) which evidence again the presence of type I intermittency at the mapping level. 


Finally, we would like to point out the relation between the form of the slow manifold and the map function. The possible explanation of the gradual change in the subthreshold oscillations and in $S$ is that each consecutive spike in the nearly periodic epoch moves slightly away from the unstable manifold at each round, just as at each iteration of the map the system moves continuously along the specific map segment.

\section{Conclusions}

In this paper we have studied in detail the regime complex self-spiking of the inertial van der Pol-FitzHugh-Nagumo system. We have shown that this dynamics arises from the intermittent locking of the large amplitude spikes to the small-amplitude chaotic background. The spike train is composed by epochs of periodic mixed-mode oscillations---which correspond to the nearly periodic windows of the chaotic attractor---, interspersed with bursts of purely chaotic motion. The duration of the residence-time on the periodic mixed-mode oscillation state exhibits a scaling law for the nearly periodic residence times with the characteristic exponent $\frac{1}{2}$. Such a behaviour corresponds to the type I intermittency route for the transition to turbulence discussed by Pomeau and Manneville \cite{pomeau}. We also demonstrated that the scaling law can be obtained directly from the number of periodic spikes in the time-series. This estimation gives the same exponent as the one obtained for the residence times.

The analysis of the recurrence map for $S$ has revealed the existence of tangency typically associated with the type I intermittent behavior. In fact, we observe a gradual increase of $S$ during each nearly periodic epoch consistent with the evolution through the narrow tunnel between the recurrence map of $S_n$ and the identity straight line. This behavior was also confirmed by the direct analysis of the Poincar\'e map of the system and showed an accurate correlation between the form of this map and the evolution of $S_n$. 

We should stress the purely deterministic character of this neuron model. Yet the output is typically composed of irregular but repeatable sequences of spikes reminiscent of those obtained in the experimental records of neuronal activity. In general the complexity of such sequences is attributed to a combination of the action of many neurons and the presence of stochastic factors in the populations. Our low dimensional model equations open the wider perspective of considering the origin of complex firing patterns as intrinsic to the dynamics of individual cells. Under this way of thinking, we might speculate on the possibility that neural coding of stimuli could arise at cellular level by a mechanism of setting a control parameter analogous to our $a$ to learn an specific firing pattern.

The relevance of this possibility to the development of spiking coding algorithms and spiking neural networks should not be underestimated. We are currently working on promising architectures that advantage of properties of our model to implement learning and computing algorithms for these type of networks. 


\end{document}